\documentclass[runningheads,dvipsnames]{llncs}
\usepackage[T1]{fontenc}
\usepackage{graphicx}
\usepackage{hyperref}

\usepackage{amsmath}
\usepackage[capitalize]{cleveref}

\usepackage[frozencache]{minted}
\newcommand{\haskell}[1]{\mintinline{haskell}{#1}}

\newcommand{\metavar}[2]{{\normalfont\color{RoyalBlue}{\textsc{#1}{#2}}}}
\newcommand{\type}[1]{{\color{purple}{#1}}}
\newcommand{\con}[1]{{\color{blue}{\mathsf{#1}}}}

\begin{document}
\title{Free Monads, Intrinsic Scoping, and Higher-Order Preunification}
\author{Nikolai Kudasov\orcidID{0000-0001-6572-7292}}
\authorrunning{Kudasov N.}
\institute{Innopolis University, Innopolis, Tatarstan Republic, Russia \\
  \email{n.kudasov@innopolis.ru}
}
\maketitle              % typeset the header of the contribution
\begin{abstract}
Type checking algorithms and theorem provers rely on unification algorithms.
In presence of type families or higher-order logic, higher-order (pre)unification (HOU)
is required. Many HOU algorithms are expressed in terms of $\lambda$-calculus
and require encodings, such as higher-order abstract syntax, which are sometimes
not comfortable to work with for language implementors.
To facilitate implementations of languages, proof assistants, and theorem provers,
we propose a novel approach based on the second-order abstract syntax of Fiore,
data types \`a la carte of Swierstra, and intrinsic scoping of Bird and Patterson.
With our approach, an object language is generated freely from a given bifunctor.
Then, given an evaluation function and making a few reasonable assumptions on it,
we derive a higher-order preunification procedure on terms in the object language.
More precisely, we apply a variant of $E$-unification for second-order syntax.
Finally, we briefly demonstrate an application of this technique
to implement type checking (with type inference) for Martin-L\"of Type Theory, a dependent type theory.
\end{abstract}

%
%

% \tableofcontents

%
\section{Introduction}

When implementing a programming language, a proof assistant, or a theorem prover,
one often relies on unification algorithms. Dealing with dependent types
and/or higher-order logics requires higher-order unification (HOU) algorithms.
Many such algorithms are available in the literature,
most influential of which are, perhaps, Huet's preunification~\cite{Huet1975}, Jensen-Pietrzykowski's full unification~\cite{JensenPietrzykowski1976} procedures,
procedures for decidable fragments~\cite{Miller1991,AbelPientka2011,LibalMiller2016} and a recent efficient implementation of full HOU~\cite{VukmirovicBentkampNummelin2021}.

HOU algorithms, such as mentioned above, are specified for a rather minimalistic version of $\lambda$-calculus.
This is often justified by appealing to higher-order abstract syntax~\cite{PfenningElliott1988} (HOAS):
any binding construction can be encoded in $\lambda$-calculus.
Unfortunately, HOAS and its variants~\cite{Chlipala2008,WashburnWeirich2008} are not always comfortable to work with as
witnessed by both language implementors~\cite{Kmett2015_SoH,Cockx2021_blog} and formalization researchers~\cite{FioreSzamozvancev2022}.

Thus, supporting higher-order (pre)unification either forces one to use HOAS
or to implement a version of a HOU algorithm from scratch for the chosen language.
This appears to be one of the main reasons for prototype implementations to omit or reduce
support for type inference and demand more explicit type annotations for the user.

Second-order abstract syntax (SOAS) and second-order equational logic~\cite{FioreHur2010}
have recently been an attractive alternative to HOAS. It has been successfully
used to generate metatheory in Agda~\cite{FioreSzamozvancev2022} and
a full $E$-unification procedure~\cite{Kudasov2023} has been developed.
Importantly, $E$-unification for SOAS is powerful enough to encode
higher-order unification problems in languages with arbitrary binders.

SOAS is freely generated from a signature,
which specifies the syntactic constructions available in the object language,
by adding variables and parametrized metavariables. Each syntactic construction
can be parametrized by a sequence of (potentially, scoped) subterms. For example,
SOAS for simply typed lambda calculus~\cite[Example~1]{FioreHur2010} is generated from a family of constructors
for all types $\type{\sigma}$ and $\type{\tau}$:
\begin{align*}
    \con{app}^{\type{\sigma},\type{\tau}} : (\type{\sigma \Rightarrow \tau}, \type{\sigma}) \to \type{\tau}
    \quad\quad\quad
    \con{abs}^{\type{\sigma},\type{\tau}} : (\type{\sigma}.\type{\tau}) \to \type{\sigma \Rightarrow \tau}
\end{align*}

Here, $\con{app}^{\type{\sigma},\type{\tau}}$ has two subterms of types\footnote{
  The double arrow ($\type{\Rightarrow}$) here corresponds to the function types in the object language,
  while single arrow ($\to$) is a part of the type signature (in the metatheory) separating
  types of subterms from the type of the resulting term for each of the syntactic constructors.
} $\type{\sigma \Rightarrow \tau}$ and $\type{\sigma}$, while
$\con{abs}^{\type{\sigma},\type{\tau}}$ has a single \emph{scoped} subterm of type $\type{\tau}$ with access to a local variable of type $\type{\sigma}$.

Although it should be possible to work with intrinsically typed SOAS
as in the example above, in this paper, we consider only untyped SOAS
since ultimately we are interested in explicit implementations of
type checking and type inference for arbitrary languages,
whose type system might not be properly embeddable in the host language.
For example, we consider the following SOAS for preterms in $\lambda\Pi$-calculus
(i.e. well-scoped but not necessarily well-typed terms):
\begin{align*}
    \con{app} : (\type{T}, \type{T}) \to \type{T}
    \quad\quad\quad
    \con{abs} : (\type{T}.\type{T}) \to \type{T}
    \quad\quad\quad
    \con{Pi}  : (\type{T}, \type{T}.\type{T}) \to \type{T}
\end{align*}

Free monads~\cite{Swierstra2008} generate (first-order) abstract syntax trees
with a monadic binding operation serving as substitution.
Following Swierstra~\cite{Swierstra2008}, we want to generate SOAS
with proper variable substitution and metavariable substitutions
from a signature provided by a user-defined algebraic data type (ADT).
To do that, we need to be able to specify and properly handle scoped terms.

For expressions with scopes (such as $\mathsf{let}$-expressions or $\lambda$-abstractions),
substitution (implemented manually or via free monads) is not safe by default since a name capture might happen.
To avoid this, de Bruijn indices \cite{deBruijn1972} are commonly used in practice.
Generalized de Bruijn indices\footnote{such as implemented in the \texttt{bound} package,
available at \url{http://hackage.haskell.org/package/bound} }
have also been used (e.g. in Epigram~\cite{McBride2004}) to keep track of scoping in types
and also to allow for the lifting entire subexpressions to optimize substitutions further.

Combining free monads with intrinsic scopes via generalized de Bruijn indices
we are able to generate abstract syntax with proper substitution operations.
For $\lambda\Pi$-calculus the following ADT describes the signature of preterms:
\begin{minted}[mathescape]{haskell}
data TermF scope term
  = LambdaF scope       -- $\lambda x. T$
  | AppF term term      -- $T_1 \; T_2$
  | PiF term scope      -- $\prod_{x : T_1} T_2(x)$
\end{minted}

Here the \haskell{scope} parameter corresponds to scoped subterms,
introducing local variable(s), and \haskell{term} corresponds
to subterms without extra scope variables. It is possible to consider
variations of our approach, supporting arbitrary indexing for bound variables
and support for intrinsic typing. However, we find the suggested setting comfortable
enough and defer variations for further work.

Since we can see the ADT above as a signature for SOAS,
we can generate syntax for the object language (with and without metavariables),
and provide higher-order preunification, adapting a version of $E$-unification for SOAS~\cite{Kudasov2023}.

\subsection{Related Work}

\paragraph{Unification and Free Monads.}

In his 2001 pearl \cite{Sheard2001},
Sheard described an efficient and modularized implementation
of single-sorted first-order unification.
Wren Romano has implemented this approach in Haskell
as the \texttt{unification-fd} library.
Romano's implementation also mixes well with Swierstra's data types \`a la carte \cite{Swierstra2008}:
terms with metavariables are constructed using free monads.

Axelsson and Vezzosi~\cite{AxelssonVezzosi2016} use data types \`a la carte
approach in their syntax for higher-order rewrite rules. However, their
implementation of capture avoiding substitution explicitly demands specific
syntax for variables, lambda abstractions, and applications, preventing
correct treatment of other potential binding constructions. Our approach
adds variables freely and does not impose any further restrictions on the syntax of the object language,
allowing arbitrary binders.

\paragraph{Second-Order Abstract Syntax.}

Fiore and Szamoszvancev~\cite{FioreSzamozvancev2022} have developed a language-formalization
framework in Agda. Their approach is based on SOAS and generates Agda code
for a grammar of types, operations of weakening and substitution, correctness properties,
and other utilities for the formalization of an equational/rewriting theory for a given language.

Makoto Hamana~\cite{Hamana2020} has developed the framework of second-order computation systems
and their algebraic semantics, laying out the foundation for the \textsf{SOL} system,
a tool to check confluence and termination of polymorphic second-order computation systems.
The \textsf{SOL} system is implemented in Haskell and relies on the quasiquotation
feature of Template Haskell to specify a second-order signature and computation rules for
a second-order computation system.

Whereas the aforementioned works are focused on the metatheory of languages,
we are interested more in the implementation of languages, and in particular,
type checkers for dependently typed languages.

\paragraph{Intrinsic Scoping.}

Maclaurin, Radul, and Paszke have introduced the Foil~\cite{MaclaurinRadulPaszke2023},
making it possible to have intrinsic scoping while maintaining the efficiency benefits of the Barendregt convention,
as implemented in the Rapier~\cite{PeytonJonesMarlow2002}, an approach to handling binders,
which is implemented, in particular, in the Glasgow Haskell Compiler.
It seems plausible that the Foil can be used instead of nested datatypes to ensure
scope safety in the approach presented in this paper, but we leave this research for future work.

\subsection{Contributions}

We propose an approach to abstract syntax
that relies on a combination of free monads and generalized de Bruijn indices.
We argue that our approach facilitates the implementation of programming languages and proof assistants\footnote{or, at the very least, prototyping of programming languages and proof assistants},
in particular, of dependently typed ones, by deriving a higher-order preunification algorithm for the object language.
Our specific contributions are the following:
\begin{enumerate}
  \item In \cref{section:free-scoped-monads}, we introduce \emph{free scoped monads}, a generic data type that serves as a basis for a family of languages with well-scoped terms.
  % We show how to work with well-scoped abstract syntax using free scoped monads.
  \item In \cref{section:reduction-a-la-carte}, we propose an approach to the implementation of term reduction that mixes well with the data types \`a la carte approach.
  \item In \cref{section:higher-order-unification}, we formulate the necessary requirements for the signature to enable higher-order preunification of terms in the object language.
    We adapt an $E$-unification procedure for second-order abstract syntax~\cite{Kudasov2023} and extract the preunification component of it.
  %\item In Section~\cref{section:type-inference}, we give a representation for typed terms based on free scoped monads, outline the bottom-up constraint based type inference, and formulate the necessary requirements on the generating bifunctor.
  \item In \cref{section:applications}, we demonstrate how our approach can be applied to implement type checking and type inference for Martin-L\"of Type Theory.
\end{enumerate}

\section{Free Monads with Intrinsic Scoping}
\label{section:free-scoped-monads}

In this section, we merge the ideas of free monads and intrinsically scoped terms
to produce free scoped monads allowing us to generate the type of well-scoped terms
with correctly defined substitution. We then add metavariables,
generating SOAS from a signature given by an algebraic data type (ADT) in Haskell.

Intrinsically well-scoped de Bruijn terms were introduced
by Bellegarde and Hook~\cite{BellegardeHook1994},
and monadic structure (substitution) for untyped terms was developed
by Bird and Patterson~\cite{BirdPatterson1999}.
Some later work has been done for typed terms~\cite{AltenkirchReus1999,AllaisAtkeyChapmanMcBrideMcKinna2018},
but those use intrinsic typing which we are not using in this paper\footnote{
  we are not relying on intrinsic typing since we would like to be able to implement
  languages with richer type systems in weaker or differently typed host languages;
  for example, we would like to implement Martin-Löf Type Theory in Haskell
}.

Skipping intrinsic typing, we are not relying on dependent types in the host language,
however our representation of abstract syntax still requires two important type system features.
First, we require \emph{nested} (also called \emph{non-uniform} or \emph{non-regular}) data types,
whose definition involves a recursive component that is different from the type being defined.
Second, we require higher-kinded types in order to parametrize the signature ADT by type constructors.
We are using Haskell as our language of implementation,
but the reader should be aware of these requirements,
if they wish to port the code to another language.

\subsection{Intrinsically Well-Scoped Terms}

Following Bird and Patterson \cite{BirdPatterson1999} we start with the following definitions.
First, we define a type constructor to extend the type of variables with one more name:

\begin{minted}{haskell}
data Inc var = Z | S var
\end{minted}

A scoped term is now a term defined in an extended context
(i.e. over the type of variables extended with (bound) variable \haskell{Z}):

\begin{minted}{haskell}
type Scope term var = term (Inc var)
\end{minted}

Note that \haskell{Scope} is a higher-kinded type since its argument \haskell{term}
is a type constructor.
As long as \haskell{term} is a \haskell{Monad}, we can perform substitution for the bound variable:
\begin{minted}{haskell}
substitute :: Monad term => term a -> Scope term a -> term a
substitute u s = s >>= \x -> case x of
    Z   -> u          -- substitute bound variable
    S y -> return y   -- keep free variable
\end{minted}

Note that intrinsic scoping here makes sure that we only substitute bound variables,
and free variables (as well as the rest of the structure) are left intact.

One could use \haskell{Scope} directly to define, for example,
the type of $\lambda\Pi$-terms, parametrized over the type of free variables:

\begin{minted}{haskell}
data Term a
  = Var a                       -- ^ Free variable.
  | App (Term a) (Term a)       -- ^ Application.
  | Lam (Scope Term a)          -- ^ Abstraction.
  | Pi (Term a) (Scope Term a)  -- ^ Function type.
\end{minted}

Assuming we have \haskell{Monad Term} instance, and equipped with \haskell{substitute},
it is straightforward to define evaluation of such terms. For example, this is how evaluation to weak head normal form (WHNF) can be implemented:
\begin{minted}{haskell}
whnf :: Term a -> Term a
whnf term = case term of
  App fun arg -> case whnf fun of
    Lam body -> whnf (substitute arg body)
    fun' -> App fun' arg
  _ -> term
\end{minted}

Compared with traditional de Bruijn indices,
relying on nested data types using \haskell{Scope} has two great advantages.
First, it is safer since ill-scoped terms are also ill-typed.
Second, programming with scopes is now more type-driven and allows for more straightforward implementations (with \haskell{substitute} being a prime example).

\subsubsection*{Binding Multiple Variables.}

It will be useful to us in \cref{subsection:terms-with-meta} to have a variation
of \haskell{Scope} that supports binding of many variables at once:
\begin{minted}{haskell}
data IncMany var
  = BoundVar Int -- an Int-indexed bound variable
  | FreeVar var  -- a free variable

-- | A scope with arbitrarily many bound variables.
type IntScope term var = term (IncMany var)
\end{minted}

Substitution for \haskell{IntScope} requires a mapping from a bound variable index to a term, but is otherwise straightforward.
\begin{minted}{haskell}
substituteMany :: Monad term => (Int -> term var) -> IntScope term var -> term var
substituteMany f s = s >>= \x ->
 case x of
   BoundVar n -> f n
   FreeVar z -> return z
\end{minted}

In this paper, we will use the regular \haskell{Scope} for the scopes in the object language and \haskell{IntScope} for metavariable substitution. % This is because it is more convenient to work with fully applied metavariables and, correspondingly, with fully abstracted scopes.

\subsection{Free Scoped Monads}

% Before generating full SOAS from a signature, we first focus on a simpler generalisation of free monads.

The use of \haskell{substitute} in the definition of \haskell{whnf} requires a \haskell{Monad} instance for \haskell{Term}.
Although we could provide it explicitly, we would rather have it for free.
One common technique to get it is to reformulate \haskell{Term} using free monads \cite{Swierstra2008}.
Unfortunately, our \haskell{Term} is used non-uniformly in its recursive definition, which is not compatible with standard free monad definitions.
So, we introduce the \emph{free scoped monad}:
\begin{minted}{haskell}
data FS t a
  = Pure a
  | Free (t (Scope (FS t) a) (FS t a))
\end{minted}

The main idea is that \haskell{t} in \haskell{FS t a} represents possible syntactic constructions
of the language (similarly to generating functor in regular free monads),
and it can explicitly mention both subterms and scopes.
The free scoped monad is a \haskell{Monad} (whenever \haskell{t} is a \haskell{Bifunctor}\footnote{Note that \haskell{Bifunctor} and some other instances can be automatically derived for a user-defined types using GHC extensions, such as \texttt{DeriveFunctor}, or Template Haskell utility functions like \texttt{deriveBifunctor} from \texttt{bifunctors} package.}),
with the bind operation (\haskell{>>=}) corresponding to the substitution.
Importantly, unlike regular free monads~\cite{Swierstra2008},
the substitution in free scoped monads respects bound variables.
\begin{minted}{haskell}
instance Bifunctor t => Monad (FS t) where
  return = Pure
  Pure x >>= f = f x
  Free t >>= f = Free (bimap ((>>= traverse f)) (>>= f) t)
\end{minted}

We now reformulate our type for untyped lambda terms, defining \haskell{TermF} to specify all syntactic constructions
and using \haskell{FS} to give us the type of terms:
\begin{minted}{haskell}
data TermF scope term
  = AppF term term
  | LamF scope

type Term a = FS TermF a
type ScopedTerm a = Scope (FS TermF) a
\end{minted}

The \texttt{PatternSynonyms} extensions helps us keep \haskell{whnf} implementation without any changes after the switch to free scoped monads representation:

\begin{minted}{haskell}
pattern Var x = Pure x
pattern Lam s = Free (LamF s)
pattern App t1 t2 = Free (AppF t1 t2)
\end{minted}

It will be useful to us sometimes to apply a transformation to all nodes,
changing from one signature to another. For that, we introduce this function:
\begin{minted}{haskell}
trans :: Bifunctor f => (forall x y. f x y -> g x y) -> FS f a -> FS g a
trans _phi (Pure x) = Pure x
trans phi (Free t) = Free (phi (bimap (trans phi) (trans phi) t))
\end{minted}

With free scoped monads, we now have the tools to generate types of well-scoped terms.
Although \haskell{FS} provides an effective mechanism to automatically get substitution for our terms,
the design as presented here has some trade-offs.
First, since we are using \haskell{Scope}, we are limiting ourselves to scopes that only introduce one bound variable.
This can be improved by using \emph{generalized de Bruijn indices} as implemented in the \texttt{bound} package.
In this paper, we will use a simplified version for the sake of clarity and brevity.
Second, the definition of \haskell{Scope} itself can be changed to reduce the number of required traversals of the syntax tree.
Again, a more elaborate version, as seen in the \texttt{bound} package, can be used instead.
Third, we could parametrize \haskell{FS} over the scope type constructor,
but that would again unnecessarily complicate the code.
Finally, we could use a different formulation of \haskell{FS},
such as a Church encoding, similar to Church-encoded free monads \cite{Voigtlander2008} for improved asymptotic complexity of substitution.

\subsection{Metavariables, SOAS, and Metavariable Substitution}
\label{subsection:terms-with-meta}

For unification, we need to add metavariables to our syntax.
To avoid unnecessary assumptions about the object language
while keeping the expressive power of higher-order unification,
we follow SOAS~\cite{FioreHur2010} and use parametrized metavariables.
Instead of embedding metavariables directly into \haskell{FS} data type,
we use data types à la carte approach~\cite{Swierstra2008}
and extend any given bifunctor \haskell{term} with metavariables.
The following datatype generates parametrized metavariables:

% We define \haskell{SOAS v term a} to be the type of second-order terms
% generated from a signature \haskell{term} with (parametrised) metavariables
% indexed in \haskell{v} and free variables indexed in \haskell{a}.
% We then define metavariable substitution.

\begin{minted}{haskell}
data MetaAppF v scope term = MetaAppF v [term]
\end{minted}

Parametrization provides independence from object language syntax
(we do not require having function application in the object language),
but it also keeps all ``dependencies'' of a metavariable bundled with it.

Following~\cite{Kudasov2023}, we write $\metavar{m}{_i}[t_1, t_2, \ldots, t_n]$ to mean application of metavariable $\metavar{m}{_i}$ to terms $t_1, t_2, \ldots, t_n$.
Note that $\metavar{m}{_i}[t_1][t_2]$ is invalid syntax, and it is not possible in general
to partially apply a metavariable.

To add metavariables to a language, we use a variant of Swierstra's operator (\haskell{:+:}).
Given signatures \haskell{f} and \haskell{g}, we can get a new signature \haskell{Sum f g} that supports constructions from both original signatures.
% Note that \haskell{Sum f g} is a \haskell{Bifunctor} whenever \haskell{f} and \haskell{g} are.
\begin{minted}{haskell}
data Sum f g scope term
  = InL (f scope term)  -- inject constructions of f
  | InR (g scope term)  -- inject constructions of g
\end{minted}

Now, we can extend any signature \haskell{t} with parametrized metavariables:
\begin{minted}{haskell}
type SOAS v t a = FS (Sum t (MetaAppF v)) a
\end{minted}
Here, \haskell{SOAS} stands for ``Second-Order Abstract Syntax'' with \haskell{v} being the type of metavariables,
\haskell{t}~--- the term signature, and \haskell{a}~--- the type of free variables.

\subsubsection*{Metavariable Substitution.}

Following SOAS~\cite{FioreHur2010}, we define substitution for parametrized metavariables
by mapping each metavariable to a scoped term, with $n$ bound variables.
In the implementation, we rely on \haskell{IntScope}, allowing arbitrarily many
bound variables (not statically checked):
\begin{minted}{haskell}
data MetaAbs t a = MetaAbs Int (IntScope (FS t) a)
\end{minted}

Here, the first component of type \haskell{Int} represents the arity of the metavariable,
which is mostly useful for pretty-printing and debugging, and is not strictly necessary for the unification algorithm.
The second component is the scoped term, with (up to) $n$ distinct bound variables used.

We represent substitution $\metavar{m}{_i}[x_1, x_2, \ldots, x_n] \mapsto t$ as
a pair of metavariable $\metavar{m}{_i}$ and the scoped term $t$, represented using \haskell{MetaAbs} for the extended language.
A simultaneous substitution~\cite[Section~2]{FioreHur2010} is represented by a list
of substitutions\footnote{Here we are using a list for simplicity, but it is also possible to use other data structures, such as \haskell{Data.HashMap} or \haskell{Data.Map}.}:
\begin{minted}{haskell}
type Subst v t a = (v, MetaAbs (Sum t (MetaAppF v)) a)
newtype Substs v t a = Substs { getSubsts :: [Subst v t a] }
\end{minted}

To apply \haskell{Substs} to a term, we merely traverse the term replacing every occurrence of \haskell{MetaAppF} that has a corresponding substitution:
\begin{minted}{haskell}
applySubsts :: (Eq v, Bifunctor t) => Substs v t a -> SOAS v t a -> SOAS v t a
applySubsts substs = go where
  go term = case term of
    Pure{} -> term -- free variables remain
    Free (InR (MetaAppF v args)) -> -- metavariables are replaced according to substs
      -- substitue metavariables in arguments
      let args' = map (applySubsts substs) args
       in case lookup v (getSubsts substs) of
            Just (MetaAbs _arity body) -> substituteMany (args' !!) body
            Nothing -> Free (InR (MetaAppF v args'))
    -- recursively traverse other syntactic constructions
    Free (InL t) -> Free (InL (bimap goScope go t))
  goScope = applySubsts (fmap S substs)
\end{minted}

% Note the use of \haskell{fmap S} that performs the lift for all substitutions when we go inside a scope.

This concludes the definition of SOAS generated from a signature provided in a form of a \haskell{Bifunctor} in Haskell:
\begin{enumerate}
    \item \haskell{SOAS v t a} is the type of second-order terms generated from \haskell{t};
    \item \haskell{Pure x} corresponds to a (free) variable $x$;
    \item \haskell{Free (InR (MetaAppF v [t1, ..., tN]))} corresponds to $\metavar{m}{}[t_1, \ldots, t_N]$
    \item \haskell{Free (InL t)} corresponds to some syntactic construction\footnote{Fiore and Hur call these \emph{operators}~\cite[Section~2]{FioreHur2010}} $\con{F}(\overline{x_1}.t_1, \ldots, \overline{x_n}.t_n)$;
    \item function \haskell{applySubsts} performs simultaneous metavariable substitution.
\end{enumerate}

\section{Term Reduction \`a la Carte}
\label{section:reduction-a-la-carte}

In this section, we organize term reduction for extensible languages following data types \`a la carte~\cite{Swierstra2008}.
The motivation is twofold. On the one hand, we want to be able to specify reduction in object languages
without having to deal with metavariables (indeed, it is natural for the reduction rules to be independent of metavariables).
On the other hand, we want to be able to easily
extend languages with new syntactic constructions (e.g. pairs and projections).
%  For an example of the latter,
% imagine extending a language with pairs and projections.
% For many languages, it is sufficient to specify syntax and reduction for pairs and projections
% in isolation and extend an arbitrary language with it.

% At this stage, we do not impose specific restrictions of the definition of ``reduction'',
% except that it should work for arbitrary terms, possibly with free variables.
% Later, when dealing with (pre)unification we will require reduction to be confluent.

In general, assuming the constructions from the two signatures are not supposed to ``interfere'' with each other,
we can define term reduction for each component independently.
To get reduction working for terms generated from signature `Sum f g`,
we need to specify reduction for `f` and `g`, however, it is important that we give that anticipating an extension\footnote{Hence the use of \haskell{Sum} in the type signature. The importance of this definition becomes clear when we consider the instance of \haskell{Redicible} for \haskell{Sum f g}}:
\begin{minted}{haskell}
class Bifunctor t => Reducible t where
  reduceL :: Reducible ext
    => t (Scope (FS (Sum t ext)) a) (FS (Sum t ext) a)
    -> FS (Sum t ext) a
\end{minted}

Here \haskell{reduceL} reduces a term of a language, generated by \haskell{t} extended with \haskell{ext},
assuming terms generated by \haskell{ext} are reducible. Using \haskell{reduceL} and commuting left and right languages,
we get \haskell{reduceR}\footnote{We can also make \haskell{reduceR} a part of the \haskell{Reducible} class.}:

\begin{minted}{haskell}
reduceR = commute . reduceL . bimap commute commute
commute = trans $ \case
  InL x -> InR x
  InR y -> InL y
\end{minted}

\subsubsection*{Empty Signature}

A particularly important language is an empty one:
\begin{minted}{haskell}
data Empty scope term   -- this data type has no constructors
\end{minted}

%The type \haskell{Empty} is vacuously \haskell{Bitraversable}.
Note that language generated by \haskell{Empty} is not actually empty: free variables are always added with \haskell{FS}.
Without any constructors \haskell{reduceL} is trivial:
\begin{minted}{haskell}
instance Reducible Empty where
  reduceL e = case e of {}
\end{minted}

We can express term reduction in an (unextended) object language as a special case of \haskell{reduceSum},
extending it with an \haskell{Empty} signature:
\begin{minted}{haskell}
reduce :: Reducible t => FS t a -> FS t a
reduce = trans removeEmpty . reduceSum . trans InL
  where removeEmpty (InL x) = x
        removeEmpty (InR e) = case e :: Empty of {}
\end{minted}

\subsubsection*{Sum of Signatures}

Combining two reducible languages with \haskell{Sum} yields a reducible language.
Here, we rely on commutativity and associativity of \haskell{Sum}:

\begin{minted}{haskell}
instance (Reducible f, Reducible g) => Reducible (Sum f g) where
  reduceL (InL t) = assoc' (reduceL (bimap assoc assoc t))
  reduceL (InR t) = from (reduceL (bimap to to t))
    where
      to = assoc . commute . assoc
      from = assoc' . commute . assoc'
\end{minted}

\subsubsection*{Reducing $\lambda\Pi$-Terms}

To adapt \haskell{whnf} to reduce terms in a language generated by \haskell{TermF} with arbitrary extension, we can introduce patterns for extended language:
\begin{minted}{haskell}
pattern LamE body = Free (InL (LamF body))
pattern AppE t1 t2 = Free (InL (AppF t1 t2))
pattern ExtE t = Free (InR t)
\end{minted}

Implementation of \haskell{whnf} is transferred almost letter for letter, with an important addition being the case when the root node belongs to the extension~--- here, we delegate reduction to the extension by using \haskell{reduceR}:

\begin{minted}{haskell}
instance Reducible TermF where
  reduceL = \case
    AppF fun arg -> case reduceSum fun of
      LamE body -> reduceSum (substitute arg body)
      fun' -> AppE fun' arg
    t -> Free (InL t)
\end{minted}

\subsubsection*{Reducing with Metavariables}

Parametrized metavariables reduce to themselves, however we can choose to reduce or keep
their parameters, yielding two possible definitions. First one leaves parameters unevaluated:

\begin{minted}{haskell}
-- "lazy" reduction (arguments remain unevaluated)
instance Reducible (MetaAppF v) where
  reduceL t = Free (InL t)
\end{minted}

The second possible implementation reduces the parameters:

\begin{minted}{haskell}
-- "strict" reduction (arguments are evaluated)
instance Reducible (MetaAppF v) where
  reduceL (MetaAppF m args) = Free (InL (MetaAppF m (map reduceSum args)))
\end{minted}

Note that even though the second implementation is ``strict'' in the object language,
using Haskell as a host language makes evaluation somewhat lazy in the sense that
actual evaluation of parameters might still be delayed. This kind of lazy evaluation
is used in some HOU algorithms~\cite{VukmirovicBentkampNummelin2021} and from now
we assume the second instance implementation for \haskell{MetaAppF}.

\section{Higher-Order Unification}
\label{section:higher-order-unification}

In this section, we describe a generic semi-decidable algorithm for single-sorted higher-order preunification.
The algorithm stops when either the terms cannot be unified,
or when the only constraints left are those between metavariables.

The algorithm is loosely based on $E$-unification for second-order abstract syntax~\cite{Kudasov2023},
with the following important differences:
\begin{enumerate}
    \item we forego the \textbf{(mutate)} rule~\cite[Definition~28]{Kudasov2023}, and instead assume that terms can be normalized (via \haskell{reduce});
    \item we combine the \textbf{(imitate)} and \textbf{(project)} rules~\cite[Definitions~24--25]{Kudasov2023} into a single rule with generalized Huet-style bindings~\cite{Kudasov2023-UNIF};
    \item we only implement preunification, leaving unsolved constraints between two metavariables, so we are not using \textbf{(eliminate)}, \textbf{(identify)}, and \textbf{(iterate)} rules;
    \item we implement \emph{unification} itself in an untyped setting, i.e. our implementation of higher-order unification does not (directly) take types of terms into account;
    technically, type information can be embedded into the terms themselves and can be used by \haskell{reduce}, but in this paper we do not make extra assumptions
    when generating Huet-style bindings and leave development of an algorithm for type-directed generalized Huet-style bindings for future work.
\end{enumerate}

To achieve such an algorithm, we impose some constraints on the signature ADT.
These constraints, formulated as type classes in Haskell,
make sure that we can traverse the freely generated syntax tree,
match individual nodes of that tree (enabling first-order unification),
and make appropriate substitutions for metavariables (enabling higher-order unification).

\subsection{Prerequisites}

The unification process involves keeping track of metavariables and their values, which is a kind of effectful computation. Traversing an abstract syntax tree and making changes to the currently known values of metavariables requires not just \haskell{Bifunctor}, but \haskell{Bitraversable} instance for the generating bifunctor. Fortunately, for most practical cases, we can derive those instances automatically with common GHC extensions or Template Haskell.

Apart from \haskell{Bitraversable}, we will require object language terms to be \haskell{Reducible}, and its syntactic constructions \haskell{Unifiable}.
For first-order unification, it would be enough to match individual syntactic constructions and perform unification by recursive matching.

For higher-order unification, we will require generalized Huet-style bindings~\cite{Kudasov2023-UNIF}.
Essentially, for a given language we need to know the following information:
\begin{enumerate}
  \item For each argument (subterm) of a syntactic construction, is there a certain shape of a term that allows further reduction?
  For example, given a term $\pi_1 \; \metavar{m}{_1}[]$ (where $\metavar{m}{_1}[]$ is a metavariable),
  we should understand that $\metavar{m}{_1}$ can be substituted by a tuple $(\metavar{m}{_2}, \metavar{m}{_3})$ (where $\metavar{m}{_2}$ and $\metavar{m}{_3}$ are fresh metavariables) to allow further reduction.
  \item For parametrized metavariables, what are possible ways to construct a term that will use
  the parameters via reduction? For example, when unifying $\metavar{m}{_1}[(t_1, t_2)]$ with $t_1$, we should be able to try the substitution $\metavar{m}{_1}[x] := \pi_1 \; \metavar{m}{_2}[x]$ to find the solution $\metavar{m}{_1}[x] := \pi_1 \; x$.
\end{enumerate}

\subsubsection*{First-Order Unification}
For first-order unification, we need to be able to match individual nodes of the syntax tree.
Similarly to Wren Romano's \texttt{unification-fd} package\footnote{In \texttt{unification-fd} and
our implementation the type of \haskell{zipMatch} is a little more complicated to allow for an optimization,
when one of the nodes omits a subterm, and we can immediately take the necessary value from the second node.
However, we decided to simplify the type here to increase readability of this paper.},
we define a type class with a single method:

\begin{minted}{haskell}
class Unifiable t where
  zipMatch :: t scope term -> t scope term -> Maybe (t (scope, scope) (term, term))
\end{minted}

The method \haskell{zipMatch} takes two nodes as inputs and returns \haskell{Nothing} when they do not match.
Otherwise, it returns a single node with subterms and subscopes paired.
For example, matching \haskell{Lam t} with \haskell{Lam u} yields \haskell{Just (Lam (t, u))},
suggesting that the unification process should now proceed by going inside the lambda and attempting to unify \haskell{t} and \haskell{u}.
\begin{minted}{haskell}
instance Unifiable TermF where
  zipMatch (AppF f1 x1) (AppF f2 x2) = Just (AppF (f1, f2) (x1, x2))
  zipMatch (LamF body1) (LamF body2) = Just (LamF (body1, body2))
  zipMatch _ _ = Nothing
\end{minted}

Implementing \haskell{Unifiable} is usually mechanical and can in fact be fully automated using
GHC's Generics (as is done in \texttt{unification-fd}) or Template Haskell.

\subsubsection*{Higher-Order Unification}
\label{subsection:higher-order-matching}

Note that, following SOAS~\cite{FioreHur2010}, we do not require the signature
to have lambda abstractions or applications as there is more than one way the user might want to introduce those.
For example, application can be defined as binary or taking a list of arguments.
Moreover, some theories have not one but several syntactic abstractions or applications
(such as $\Pi$-types and extensions types in Riehl and Shulman's type theory with shapes \cite{RiehlShulman2017},
or $\mu$-abstraction in Parigot's $\lambda\mu$-calculus~\cite{Parigot1992}).

So instead of forcing syntax onto the user, we instead ask them to provide a basic mechanism
for generating valid structural guesses (generalized Huet-style bindings) for metavariables:
\begin{minted}{haskell}
-- | Placeholder for a subterm that may or may not contain the head.
data IsHead = HasHead | NoHead

class Unifiable t => HigherOrderUnifiable t where
  guessMetas :: t scope term -> t (scope, [t () ()]) (term, [t () ()])
  shapes :: [t IsHead IsHead]
\end{minted}

The role of \haskell{guessMetas} is to provide a list of valid partial guesses for each subterm and subscope in a given node of the syntax tree. For example, given a term $\metavar{m}{_1} \; \metavar{m}{_2}$ where $\metavar{m}{_1}$ and $\metavar{m}{_2}$ are metavariables we can guess that $\metavar{m}{_1}$ is a function and so should be unified with a term $\lambda \metavar{m}{_3}$, where $\metavar{m}{_3}$ is a fresh metavariable. On the other hand, we do not have any information that would allow us to guess the structure of $\metavar{m}{_2}$.
Each returned guess for a particular subterm or subscope has type \haskell{t () ()}, which simply provides the general shape of the guess (e.g., that it should be a lambda abstraction). The unification algorithm will then replace each \haskell{()} in a guess with a fresh metavariable and continue the unification process.

Implementing this for \haskell{TermF} we get the following:
\begin{minted}{haskell}
instance HigherOrderUnifiable TermF where
  guessMetas term = case term of
    AppF f arg -> AppF (f, [LamF ()]) (arg, [])
    _ -> bimap (,[]) (,[]) term

  shapes = [AppF HasHead NoHead]
\end{minted}

Note that the type of \haskell{guessMetas} implies that a guess is based on a single syntactic construction (i.e. it cannot match a complex pattern). It also yields just one syntactic construction per guess (with placeholders for fresh metavariables).

For many type theories, one only needs to identify introduction-elimination pairs to implement \haskell{HigherOrderUnifiable}. Given an instance of \haskell{Reducible}, one can go over all possible combinations of syntactic constructions to figure out this information automatically, either using Template Haskell or GHC Generics.

\subsection{Constraints}
\label{subsection:constraints}

A constraint is essentially a pair of terms with metavariables that we would like to unify.
Importantly, the same metavariable can be used with different parameters and in different scopes.
This means that a metavariable substitution cannot depend on the bound variables (otherwise they may ``leak'').

Consider constraint involving $\lambda$-abstraction: $(\lambda f. \lambda x. \metavar{m}{_1}[f \; x]) \equiv (\lambda f. \lambda x. f \; x)$. Going under $\lambda$-abstration in both terms might reduce the original constraint to $\metavar{m}{_1}[f \; x] \equiv f \; x$. However, treating $f$ and $x$ now as free variables is incorrect as this constraint can be satisfied with two different substitutions: $\metavar{m}{_1}[z] \mapsto f \; x$ (leaks $f$ and $x$) and $\metavar{m}{_1}[z] \mapsto z$ (correct).

Since bound variables are not allowed to leak into solutions for unification problem, an appropriate simplification of the original constraint $(\lambda f. \lambda x. \metavar{m}{_1}[f \; x]) \equiv (\lambda f. \lambda x. f \; x)$ should look like $\forall f, x. (\metavar{m}{_1}[f \; x] \equiv f \; x)$. Here, $f$ and $x$ remain bound and are easy to avoid when generating substitutions for $\metavar{m}{_1}$. Fortunately, we can leverage \haskell{Scope} to manage $\forall$ quantifier and represent constraints properly:
\begin{minted}{haskell}
data Constraint v t a
  = SOAS v t a :=: SOAS v t a
  | ForAll (Scope (Constraint v t) a)
\end{minted}

The infix constructor \haskell{(:=:)} is used to construct a simple constraint with two terms. The constructor \haskell{ForAll} uses \haskell{Scope} to construct a scoped constraint. This representation does not solve the problem of leaking bound variables completely on its own, but it makes the compiler reject implementations that do not account for bound variables, as those substitutions will be impossible to lift outside of scopes.

\subsection{Preunification Algorithm}

In this section, we describe an algorithm for single-sorted preunification.
The algorithm relies on a backtracking-capable environment and the ability to generate fresh metavariables.
In Haskell, we manage those capabilities via type classes \haskell{MonadPlus} and \haskell{MonadFresh}:
\begin{minted}{haskell}
class Monad m => MonadFresh v m | m -> v where
  freshMeta :: m v
\end{minted}

The main idea of the algorithm is straightforward:
\begin{enumerate}
  \item starting with a collection of constraints,
  \item attempt to simplify them into smaller constraints by using term reduction and structural guesses for metavariables, producing some flex-flex and flex-rigid constraints;
  \item then take any flex-rigid constraint that could not be simplified further and try to solve it;
  \item if cannot be solved~--- backtrack; otherwise~--- apply solution (substitution) to the rest of the constraints and
  \item repeat until all flex-rigid constraints are resolved.
\end{enumerate}

\subsubsection*{Simplifying Constraints}

Simplification of a single constraint consists of three steps:
\begin{enumerate}
  \item Terms are reduced using \haskell{reduce}.
  \item Each term is traversed to see if any metavariables can be substituted using one of the structural guesses using \haskell{guessMetas}. If there are any potential substitutions, we apply them to both terms and repeat from step~1.
  \item Finally, we \haskell{zipMatch} the two terms to break down constraint into a collection of smaller constraints.
\end{enumerate}

Given a collection of constraints, we perform simplification on each of them recursively, accumulating and applying generated substitutions from \haskell{guessMeta}, until we end up with a collection of constraints that cannot be simplified any further.

Simplified constraints are expected to be
\begin{itemize}
  \item of the form $\forall y_1 \ldots y_m. (M_i[t_1, \ldots, t_n] \equiv t)$, where $t$ is not a metavariable application; these are called \emph{flex-rigid} constraints;
  \item or of the form $\forall y_1 \ldots y_m. (M_i[t^i_1, \ldots, t^i_n] \equiv M_j[t^j_1, \ldots, t^j_k])$; these are called \emph{flex-flex} constraints.
\end{itemize}

The third potential type of constraints, where both sides are not metavariable applications, are called \emph{rigid-rigid} constraints. These constraints are guaranteed to be simplified in step~3 with \haskell{zipMatch}. Indeed, \haskell{zipMatch} either returns \haskell{Nothing} (which means that two nodes do not match), or it pairs syntactic subtrees to match recursively, ensuring structural recursion.

\subsubsection*{Extracting Head of a Term}

If, according to \haskell{guessMetas}, there is a structural guess for some subterm position of a syntactic construction, we call a subterm in that position a \emph{head subterm}. We say that a term $h$ is a \emph{head} of a term $t$ if it is a head subterm of $t$ or if it is a head of any head subterm of $t$. For example, the term $\lambda z. f z$ is the head subterm of the term $\pi_1\;((\lambda z. f z)\;x\;(\pi_2\;y))$.

\subsubsection*{Solving Flex-Rigid Constraints}

Preunification algorithm starts off with a list of constraints, reduces \emph{rigid-rigid} constraints and solves \emph{flex-rigid} constraints, leaving only \emph{flex-flex} constraints to be dealt with by the user.

To solve a flex-rigid constraint $\forall y_1 \ldots y_m. (M_i[t_1, \ldots, t_n] \equiv t)$,
the algorithm goes through a sequence of candidate solutions. Each candidate solution is of the form $M_i[x_1, \ldots, x_n] \mapsto T$, where $T$ is one of the following:
\begin{itemize}
  \item the head of $t$, where each variable bound by $\forall$ is replaced with a fresh metavariable application $M_k[x_1, \ldots, x_n]$;
  \item a bound variable of $M_i$: $x_j$;
  \item a candidate shape (one of \haskell{shapes}), where each \haskell{HasHead} position is filled with $T'$ and \haskell{NoHead} position is filled with a fresh metavariable application $M_k[x_1, \ldots, x_n]$.
\end{itemize}

The entire algorithm is packed into a single function
with the following type signature:

\begin{minted}{haskell}
unify
  :: ( HigherOrderUnifiable t, Reducible t
     , MonadPlus m, MonadLogic m, MonadFresh v m
     , Eq a, Eq v )
  => Substs v t a
  -> [Constraint v t a]
  -> m ([Constraint v t a], Substs v t a)
\end{minted}

\section{Applications}
\label{section:applications}

In this section, we see the application of our approach to implementation of
type inference for a couple of type theories. The implementation is available
as part of version 0.1.0 of the proof assistant \textsc{Rzk}\footnote{see relevant modules in \url{https://github.com/rzk-lang/rzk/tree/v0.1.0/rzk/src/Rzk/Free}}
and contains the following relevant modules:

\begin{enumerate}
  \item \haskell{module Rzk.Free.Syntax.FreeScoped} introduces the free scoped monads, \haskell{Sum}, and utility functions
  as described in \cref{section:free-scoped-monads};
  \item \haskell{module Rzk.Free.Syntax.FreeScoped.Unification2}
  implements higher-order preunification as described in \cref{section:higher-order-unification};
  \item \haskell{module Rzk.Free.Syntax.FreeScoped.TypeCheck}
  implements type checking and type inference algorithms based on higher-order preunification;
  \item \haskell{module Rzk.Free.Syntax.Example.ULC} implements untyped $\lambda$-calculus
  with higher-order unification;
  \item \haskell{module Rzk.Free.Syntax.Example.STLC} implements a version of
  simply typed $\lambda$-calculus (STLC) with type inference via higher-order unification;
  this version differs from the standard STLC by allowing computation at the type level;
  \item \haskell{module Rzk.Free.Syntax.Example.MLTT} contains the implementation
  of intensional Martin-Löf dependent type theory with type inference.
\end{enumerate}

The type inference algorithm follows the general structure of constraint-based
typechecking, where higher-order preunification is used to resolve constraints.
For the typechecking preunification usually suffices, since flex-flex constraints
correspond to ambiguous typing which normally is considered a type error.
The details of type inference algorithm can be found in \cref{section:type-inference}.

We now outline the key moments in the implementation of Martin-Löf type theory,
more details on this and the implementation of simply typed $\lambda$-calculus
can be found in \cref{section:examples} and in the corresponding implementation files.

\subsection{Typed terms}

Many implementors define a single type in the host language for both terms and types in the object language \cite{LoehMcBrideSwierstra2010,CoquandKinoshitaNordstromTakeyama2009}. This means that typing is treated as a relation between a term and another term. We take a similar approach, annotating every node in the syntax tree with another term, which represents the type and has annotations of its own. To achieve that, we extend the object language by modifying the generating bifunctor:
\begin{minted}{haskell}
-- | Extending a type of types with universe.
data WithUniverse ty = BigUniverse | SomeType ty

data TyF t scope term = TyF
  { termF :: t scope term
  , typeF :: WithUniverse term
  }

-- | A typed term generated from t.
type TFS t a = FS (TyF t) a
\end{minted}

We use the type \haskell{WithUniverse (TFS t a)} for type annotations, meaning that type terms themselves have type annotations. The recursive annotation stops either at variables, or at \haskell{BigUniverse}, which is an explicit universe type $\mathcal{U_\infty}$. Consider term $\lambda x. f\;x$. Adding type annotations (written $t : T$) according to \haskell{TyF} would produce the following typed term (here we assume the object language also has its own universe type $\mathcal{U}$, and $f$, $A$, $B$ are free variables):
\begin{align}
  &\lambda x. f \; x \tag{untyped term}\\
  &(\lambda x. (f\;x : B)) : (A \to B : (\mathcal{U} : \mathcal{U_\infty})) \tag{typed term}
\end{align}

Since we have modified the type of nodes in the syntax tree, with \haskell{TFS t a}, we have type annotation \emph{for every subterm except variables}. This makes it easy to extract types of subterms when necessary without the need to repeatedly infer types.

% With type inference, we also need to take into account meta variables. Extending typed terms with meta variables yields the following type:
% \begin{minted}{haskell}
% type TSOAS v t a = SOAS v (TyF t) a
% \end{minted}

% Note that the universe $\mathcal{U_\infty}$ is not available as a term, it can only be used in the type position. This, in particular, means that no variable or meta variable can be instantiated with $\mathcal{U_\infty}$.

\subsection{Typing syntactic constructions}

To perform type inference for any given language, it is enough to know how to perform a single step: given types of parts for single syntactic construction, compute the type of the whole. An important implementation detail is to provide not just the types of the parts, but an actual computation context for that type. In other words, instead of \haskell{TFS v t a} we will have \haskell{m (TFS v t a)} where \haskell{m} is some typechecking monad. This is done to give the implementor of a particular language more control over typechecking and constraint resolution:

\begin{minted}{haskell}
class Inferable t where
  inferF :: MonadTypecheck v t a m
    => t (m (Scope (TFS v t) a)) (m ((TFS v t a) a))
    -> m (t (Scope (TFS v t) a) ((TFS v t a) a))
\end{minted}

Once we know how to perform a single step of type inference, all we need to do is traverse the entire term:

\begin{minted}{haskell}
infer :: (Inferable t, MonadTypecheck v t a m) => FS t a -> m (TFS v t a)
infer term = case term of
  Var x -> do
    addKnownFreeVar x
    return (Var x)
  Free t -> do
    ty <- Free <$> inferTypeFor (bimap inferScope infer t)
    clarifyTypedTerm ty
\end{minted}

Here, \haskell{addKnownFreeVar} adds the free variable to the \haskell{TypeInfo} state with a fresh type meta variable, if it is the first time this variable is encountered. As performing inference for a single syntactic construction may result in new meta variable substitutions, we need to apply them across known type information and, perhaps, simplify the inferred typed term. For that we use \haskell{clarifyTypedTerm}, which has a straightforward implementation that we omit here.

\subsection{Martin-L\"of Type Theory}

Let us now apply the approach to an actual dependent type theory~--- intensional Martin-L\"of Type Theory (MLTT). We start with a generating bifunctor:
\begin{minted}[mathescape]{haskell}
data TermF scope term
  = UniverseF         -- $\text{Universe type:\;\;} \mathcal{U}$
  | PiF term scope    -- $\text{Dependent product\;\;} \Pi_{x:T_1} T_2$
  | LamF scope        -- $\text{Abstraction:\;\;} \lambda x. T_2$
  | AppF term term    -- $\text{Application:\;\;} (T_1\;T_2)$
  | SigmaF term scope -- $\text{Dependent sum\;\;} \Sigma_{x : T_1} T_2$
  | PairF term term   -- $\text{Pair:\;\;} \langle T_1, T_2 \rangle$
  | FirstF term       -- $\text{First projection:\;\;} \pi_1\;T$
  | SecondF term      -- $\text{Second projection:\;\;} \pi_2\;T$
  | IdTypeF term term -- $\text{Identity type:\;\;} x = y$
  | ReflF term        -- $\text{Reflexivity:\;\;} \mathsf{refl}_T$
  | JF term term term term term term
    -- ^ $\text{Identity type eliminator:\;\;} \mathsf{J}(A, a, C, d, x, p)$

-- | An MLTT term with free variables in a.
type Term a = FS TermF a
\end{minted}

We note a couple of details about this particular presentation of MLTT:
\begin{enumerate}
  \item We omit type annotations for the bound variable of $\lambda$-abstraction.
  \item Both types and terms are generated with \haskell{TermF}.
\end{enumerate}

% Note that in this representation we chose to not have any type annotations for
% bound variables in abstraction and for the type of terms in the identity type
% or $\mathsf{refl}_t$. We also note that it might be possible to avoid
% the term $t$ in the annotation for $\mathsf{refl}_t$ as well,
% since the term $t$ is present in the type $t = t$ of $\mathsf{refl}$ and
% can be inferred in principle.

In this particular implementation we use a single universe type
and assume type-in-type: $\mathcal{U} : \mathcal{U}$.
It is possible to introduce a hierarchy of universes
$\mathcal{U}_0 : \mathcal{U}_1 : \mathcal{U}_2 : \ldots$
instead by using \haskell{UniverseF Natural} constructor.

Next step is to introduce helpful pattern synonyms. We will immediately work with typed terms, so we only create patterns for those. We remind that these can be automatically generated using Template Haskell:

\begin{minted}{haskell}
pattern Typed ty t = Free (InL (TyF t ty))
pattern UniverseT ty     = Typed ty UniverseF
pattern PiT ty t1 t2     = Typed ty (PiF t1 t2)
pattern LamT ty body     = Typed ty (LamF body)
pattern AppT ty t1 t2    = Typed ty (AppF t1 t2)
...
pattern JT ty t1 t2 t3 t3 t4 t5 t6 = Typed ty (JF t1 t2 t3 t4 t5 t6)
\end{minted}

Implementing WHNF reduction for MLTT is straightforward, we will focus here only on the case of $\mathsf{J}$-eliminator:
\begin{minted}{haskell}
instance Reducible TermF where
  reduceL = \case
    JF tA a tC d x p ->
      case reduce p of
        Refl{} -> reduce d
        p'     -> J tA a tC d x p'
    ...
\end{minted}

For \haskell{Unifiable} and \haskell{HigherOrderUnifiable} we also rely on a mechanical or automatic derivation and so omit it here to save space. Finally, we define inference for individual syntactic constructions:
\begin{minted}{haskell}
instance Inferable TermF
  inferF term = case term of
\end{minted}

To avoid infinite type annotations, we set the type of universe to be $\mathcal{U}_\infty$:

\begin{minted}{haskell}
    UniverseF -> pure (TyF UniverseF BigUniverse)
\end{minted}

Inferring types for $\Pi$-types and $\Sigma$-types involves dependent type checking. Given term $\Pi_{x:A} B$, where $B$ is a subterm that may refer to $x$, we have to check that both $A : \mathcal{U}$ and $B : \mathcal{U}$. Note that since $B$ is in the scope, its inferred type, by default, might also be dependent on $x$. For example, in the term $\Pi_{x : A} \mathsf{refl}_x$ the algorithm would infer that $\mathsf{refl}_{x}$ has type $x = x$, which captures the variable $x$. To make sure the body of a $\Pi$-type is always a type, we need to unify it with $\mathcal{U}$. But for that we also need to make sure it is not dependent, so we use \haskell{nonDep}:

\begin{minted}{haskell}
    PiF inferA inferB -> do
      a <- inferA >>= shouldHaveType (UniverseT BigUniverse)
      typeOfA <- typeOf a
      b <- inScope typeOfA inferB
      typeOfB <- typeOfScope typeOfA b >>= nonDep
      typeOfB `shouldHaveType` UniverseT BigUniverse
      pure (TyF (PiF a b) (UniverseT BigUniverse))
\end{minted}

Inferring the type for a dependent $\lambda$-abstraction is relatively straightforward.
We generate a fresh type meta variable for the argument and infer the type of the body.
In general, we should check that the inferred type is indeed a type, as many type theories, such as cubical type theory, have multiple universes. That said, in pure MLTT we can omit this check.

\begin{minted}{haskell}
    LamF inferBody -> do
      a <- freshTypeMetaVar
      typedBody <- inScope a inferBody
      b <- typeOfScope a typedBody
      typeOfScope a b >>= nonDep
        >>= shouldHaveType (UniverseT BigUniverse)
      pure $ TyF
        (LamF typedBody)
        (SomeType (PiT (UniverseT BigUniverse) a b))
\end{minted}

The rest of syntactic constructors is fairly straightforward to handle similarly.
Completing \haskell{Inferable} brings dependent type inference to MLTT.

\section{Conclusion and Future Work}

We have presented an approach to abstract syntax representation with free scoped monads
and demonstrated its effectiveness for the implementation of Martin-L\"of Type Theory.
Our example demonstrates that the approach does not require the user to have
a deep understanding of higher-order unification to enable type inference for their language.

We have also devised a few directions for future work.
First, we would like to extend to full higher-order unification or, better yet,
full $E$-unification for second-order abstract syntax~\cite{Kudasov2023}.
Implementing generic $E$-unification for second-order abstract syntax
would be instrumental to implementing proof assistants for type theories
with non-trivial or extensible definitional equalities.
In particular, we think this might be useful for the implementation of extension types
in Riehl and Shulman's type theory for synthetic $\infty$-categories \cite{RiehlShulman2017}.

Second, we should make higher-order unification more efficient
by optimizing the representation of free scoped monads,
taking into account the types of unified terms,
and recognizing efficient/decidable fragments of unification problems
with oracles as in the work of Vukmirovic, Bentkamp, and Nummelin~\cite{VukmirovicBentkampNummelin2021}.

\subsubsection{Acknowledgements}
I am grateful to Benedikt Ahrens and Daniel de Carvalho for their invaluable feedback
throughout my early work towards the implementation of typecheckers for dependently typed languages.
I thank Oksana Zhirosh, Ruslan Saduov, and Benedikt Ahrens for proofreading earlier versions of this paper.
I also thank the anonymous reviewer of TFP 2024 for an improvement suggestion for \haskell{Reducible}.
% I thank the anonymous reviewers of PLDI 2022, ICFP 2022, and IFL 2023 for their detailed comments
% and suggestions on the earlier versions of this paper.

\bibliographystyle{splncs04}
\bibliography{ms}

\appendix

\section{Type inference}
\label{section:type-inference}

In this section, we describe a generic type inference algorithm for languages generated using free scoped monads. As we follow a common bottom-up constraint based type inference approach, similar to existing implementations, we do not go into all the details, and instead point out the most significant definitions and aspects. % Full implementation can be found at \todo{link to implementation?}.

\subsection*{Typed terms}

Many implementors define a single type in the host language for both terms and types in the object language \cite{LoehMcBrideSwierstra2010,CoquandKinoshitaNordstromTakeyama2009}. This means that typing is treated as a relation between a term and another term. We take a similar approach, annotating every node in the syntax tree with another term, which represents the type and has annotations of its own. To achieve that, we extend the object language by modifying the generating bifunctor:
\begin{minted}{haskell}
-- | Extending a type of types with universe.
data WithUniverse ty = Universe | SomeType ty

data TyF t scope term = TyF
  { termF :: t scope term
  , typeF :: WithUniverse term
  }

-- | A typed term generated from t.
type TFS t a = FS (TyF t) a
\end{minted}

We use the type \haskell{WithUniverse (TFS t a)} for type annotations, meaning that type terms themselves have type annotations. The recursive annotation stops either at variables, or at \haskell{Universe}, which is an explicit universe type $\mathcal{U_\infty}$. Consider term $\lambda x. f\;x$. Adding type annotations (written $t : T$) according to \haskell{TyF} would produce the following typed term (here we assume the object language also has its own universe type $\mathcal{U}$, and $f$, $A$, $B$ are free variables):
\begin{align}
  &\lambda x. f \; x \tag{untyped term}\\
  &(\lambda x. (f\;x : B)) : (A \to B : (\mathcal{U} : \mathcal{U_\infty})) \tag{typed term}
\end{align}

Since we have modified the type of nodes in the syntax tree, with \haskell{TFS t a}, we have type annotation \emph{for every subterm except variables}. This makes it easy to extract types of subterms when necessary without the need to repeatedly infer types.

With type inference, we also need to take into account meta variables. Extending typed terms with meta variables yields the following type:
\begin{minted}{haskell}
type TSOAS v t a = SOAS v (TyF t) a
\end{minted}

Note that the universe $\mathcal{U_\infty}$ is not available as a term, it can only be used in the type position. This, in particular, means that no variable or meta variable can be instantiated with $\mathcal{U_\infty}$.

\subsection*{Type checking context}

We implement bottom-up type inference and keep track of currently available type information.
As we traverse a given term and solve arising constraints, this information is updated. In this subsection, we explain what kind of type information we need to store and how we mix stateful computations with backtracking.

%\subsection*{Representing current type information}

At any given moment in the algorithm, we are considering a subterm, possibly located inside several scopes. For the type inference algorithm, we translate nested data types with \haskell{Inc a} into \haskell{IncMany a}, effectively merging individual scopes into one.
\begin{enumerate}
  \item Known types of free variables. Types of free variables cannot depend on any bound variables, so for each free variable we store its type as \haskell{TSOAS v t a}.
  \item Known types of meta variables. Meta variables are global and, similarly to free variables, cannot depend on bound variables. So for each meta variable we store its type as \haskell{TSOAS v t a}.
  \item Known types of bound variables. Types of bound variables may depend on previously introduced bound variables. We store these as a list of types:\\
    \haskell{[TSOAS v t (IncMany a)]}.
  \item Known substitutions for meta variables. This is the same as in the unification algorithm with only difference being that substitutions are happening for typed terms: \haskell{Substs v (TyF t) a}.
  \item Leftover unification constraints. Again, similar to the unification algorithm, each constraint has type \\ \haskell{Constraint v (TyF t) a}.
  \item A stream of fresh meta variable identifiers.
\end{enumerate}

All of this is collected into a single data type:
\begin{minted}{haskell}
data TypeInfo v t a = TypeInfo
  { typesOfFreeVars  :: [(a, TSOAS v t a)]
  , typesOfBoundVars :: [TSOAS v t (IncMany a)]
  , typesOfMetaVars  :: [(v, TSOAS v t a)]
  , metaVarSubsts    :: Substs v (TyF t) a
  , constraints      :: [Constraint v (TyF t) a]
  , freshMetaVars    :: [v]
  }
\end{minted}

%\subsection*{The type checking monad}

To go through candidate substitutions for meta variables we rely on \haskell{MonadPlus} type class.
Moreover, we require the monad to obey the left distributive law, as it is essential for backtracking:
\begin{minted}{haskell}
  mplus a b >>= f  =  mplus (a >>= f) (b >>= f)
\end{minted}

A well-established implementation for backtracking is Kiselyov, et al.'s \haskell{LogicT} monad transformer \cite{Kiselyov2005}.
To deal with state and possible type errors we use \haskell{StateT} and \haskell{ExceptT} transformers \cite{LiangHudakJones1995} correspondingly.

Unfortunately, \haskell{StateT} does not mix well with non-deterministic nature of \haskell{LogicT}.
In particular, neither \haskell{StateT s (LogicT m)} nor \haskell{LogicT (StateT s m)} support the left distributive law of \haskell{MonadPlus}.
A common workaround is to make the state itself nondeterministic.
More specifically, we use the following data type to represent stateful computation with backtracking:
% -- | A monad with stateful computation,
% -- backtracking and error handling.
\begin{minted}{haskell}
newtype SEL s e x = SEL
  { runSEL :: StateT (Logic s) (ExceptT e Logic) x }
\end{minted}

Using \haskell{Logic s} as the type of state allows for \haskell{MonadPlus} instance that supports left distributive law:
\begin{minted}{haskell}
instance MonadPlus (SEL s e) where
  mzero = SEL (lift (lift mzero))
  mplus (SEL l) (SEL r) = SEL $ do
    states <- get
    (x, s') <- lift $ ExceptT $ mplus
      (runExceptT (runStateT l states))
      (runExceptT (runStateT r states))
    put s'
    return x
\end{minted}

We note that it is also possible to use \haskell{interleave} instead of \haskell{mplus}
to force interleaving of branches in the search space.
But it is also possible to leave more control on the user side,
deriving \haskell{MonadLogic} instance.
It is also fairly straightforward to implement \haskell{MonadState s} instance for \haskell{SEL s e}.
With all those instances in place, the monad for type checking and type inference becomes merely a special case of \haskell{SEL}:

% -- | A monad for type checking terms,
% -- generated by bifunctor t
% -- with meta variables in v
% -- and free variables in a.
\begin{minted}{haskell}
type TypeCheck v t a =
  SEL (TypeInfo v t a)
      (TypeError (TSOAS v t a))
\end{minted}

\subsection*{Typing syntactic constructions}

To perform type inference for any given language, it is enough to know how to perform a single step: given types of parts for single syntactic construction, compute the type of the whole. An important implementation detail is to provide not just the types of the parts, but an actual computation context for that type. In other words, instead of \haskell{TFS v t a} we will have \haskell{m (TFS v t a)} where \haskell{m} is some typechecking monad. This is done to give the implementor of a particular language more control over typechecking and constraint resolution:

\begin{minted}{haskell}
class Inferable ty t where
  inferF :: MonadTypecheck v t a m
    => t (m (Scope (TFS v t) a))
         (m ((TFS v t a) a))
    -> m (t (Scope (TFS v t) a)
            ((TFS v t a) a))
\end{minted}

Once we know how to perform a single step of type inference, all we need to do is traverse the entire term:

\begin{minted}{haskell}
infer :: (Inferable t, MonadTypecheck v t a m)
      => FS t a -> m (TFS v t a)
infer term = case term of
  Var x -> do
    addKnownFreeVar x
    return (Var x)
  Free t -> do
    ty <- Free <$>
      inferTypeFor (bimap inferScope infer t)
    clarifyTypedTerm ty
\end{minted}

Here, \haskell{addKnownFreeVar} adds the free variable to the \haskell{TypeInfo} state with a fresh type meta variable, if it is the first time this variable is encountered. As performing inference for a single syntactic construction may result in new meta variable substitutions, we need to apply them across known type information and, perhaps, simplify the inferred typed term. For that we use \haskell{clarifyTypedTerm}, which has a straightforward implementation that we omit here.

\subsection*{Unifying types}

Type checking in our implementation is merely a combination of type inference and unification:

\begin{minted}{haskell}
typecheck term ty = infer term >>= shouldHaveType ty

shouldHaveType term expected = do
  actual <- typeOf term
  unifyWithExpected actual expected
\end{minted}

Here, \haskell{typeOf} is a helper that either extracts the type annotation directly from \haskell{TyF}, or, when the term is a variable, extracts it from the \haskell{typesOfFreeVars} in current type information state.

For the unification, we take all known substitutions and constraints and run the pre-unification algorithm with \haskell{unify}, updating the type information and refining the types:

\begin{minted}{haskell}
unifyWithExpected actual expected = do
  substs <- gets metaVarSubsts
  cs <- gets constraints
  (cs', substs') <-
    unify substs ((actual :~: expected) : cs)
  modify (\info -> info
    { metaVarSubsts = substs'
    , constraints = cs'
    })
  clarifyTypedTerm actual
\end{minted}

\subsection*{Fresh type meta variables}

Whenever a fresh type meta variable is created, we take into account all the bound variables present in scope. In other words, we generate a meta variable application with all bound variables as arguments: $\metavar{m}{}[x_1, \ldots, x_n]$. Note that we could also add all free variables, but in practice that is rarely wanted.

\begin{minted}{haskell}
freshTypeMetaVar
  :: MonadTypecheck v t a m => m (TFS v t a)
\end{minted}

% Note that here, we make all fresh type meta variables have type \haskell{Universe}.  It is also possible to provide a more general version, where the user provide the desired type for the meta variable. Such generalization is useful, for example, when object language has to support multiple universes, when meta variables are generated for type families, or even when meta variables are used to infer terms.

\subsection*{Entering and exiting scopes}

To infer types inside scopes we introduce a couple of helpers. First, \haskell{inScope} one adds information about the type of a bound variable to the current state before running given computation in scope, then it exits the scope, removing information about the bound variable.
Second, we introduce a helper, similar to \haskell{typeOf}, that figures out types for scopes.

\begin{minted}{haskell}
inScope :: MonadTypecheck v t a m
  => TFS v t a -> m r -> m r

typeOfScope :: MonadTypecheck v t a m
  => TFS v t a
  -> Scope (TFS v t) a -> m (Scope (TFS v t) a)
\end{minted}

With these helpers we are finally ready to consider implementations of specific type theories.

\section{Examples}
\label{section:examples}

\subsection{Simply typed lambda calculus}

Here we apply our approach to an implementation of simply typed lambda calculus (STLC) with pairs. We start with a generating bifunctor:
\begin{minted}[mathescape]{haskell}
data TermF scope term
  = FunF term term          -- $\text{Function type:\;\;} T_1 \to T_2$
  | LamF (Maybe term) scope -- $\text{Abstraction:\;\;} \lambda (x : T_1). T_2$
  | AppF term term          -- $\text{Application:\;\;} (T_1\;T_2)$
  | PairTyF term term       -- $\text{Pair type:\;\;} \langle T_1, T_2 \rangle$
  | PairF term term         -- $\text{Pair:\;\;} \langle T_1, T_2 \rangle$
  | FirstF term             -- $\text{First projection:\;\;} \pi_1\;T$
  | SecondF term            -- $\text{Second projection:\;\;} \pi_2\;T$

-- | An STLC term with free variables in a.
type Term a = FS TermF a
\end{minted}

We note a couple of details about this particular presentation of STLC:
\begin{enumerate}
  \item We do not have an explicit universe type, as it is introduced automatically with \haskell{TyF}.
  \item We have an optional type annotation for the bound variable of $\lambda$-abstraction.
  The annotation is optional to illustrate how our type inference mixes with type annotations provided by the user.
  \item Both types and terms are generated with \haskell{TermF}.
\end{enumerate}

Next step is to introduce helpful pattern synonyms. We will immediately work with typed terms, so we only create patterns for those. We remind that these can be automatically generated using Template Haskell:

\begin{minted}{haskell}
pattern Typed ty t = Free (InL (TyF t ty))
pattern FunT ty t1 t2    = Typed ty (FunF t1 t2)
pattern LamT ty body     = Typed ty (LamF body)
pattern AppT ty t1 t2    = Typed ty (AppF t1 t2)
...
\end{minted}

Using these patterns we implement WHNF reduction for typed STLC terms:
\begin{minted}[mathescape]{haskell}
instance Reducible TermF where
  reduce = \case
    FirstF t -> case reduce t of
      Pair f _ -> reduce f
      t' -> First t'
    SecondT t -> case reduce t of
      Pair _ s -> reduce s
      t' -> Second t'
    App fun arg -> case reduce fun of
      Lam body -> reduce (substitute arg body)
      fun' -> App fun' arg
    term -> Free (InL term)
\end{minted}

First-order unification requires \haskell{Unifiable} instance, which has a straightforward implementation. Here we show the less trivial case for \haskell{LamF}:
\begin{minted}{haskell}
instance Unifiable TermF where
  zipMatch (LamF ty1 body1) (LamF ty2 body2) = Just (LamF ty (body1, body2))
    where
      ty = case (ty1, ty2) of
        (Nothing, Nothing) -> Nothing
        (Just t1, Just t2) -> Just (t1, t2)
        (Just t, Nothing) -> Just (t, t)
        (Nothing, Just t) -> Just (t, t)
  ...
  zipMatch _ _ = Nothing
\end{minted}

\begin{remark}
Since the type annotation for the bound variable is optional, it is possible that during unification we have the annotation on the left but not on the right, or vice versa. In this case we intend to keep the type annotation, so we pair it with itself. A more refined version of \haskell{Unifiable} type class, such as in Wren Romano's \texttt{unification-fd}, could handle this more gracefully, avoiding generating the unnecessary constraint of the form $t \equiv t$.
\end{remark}

Next, for higher-order unification we need to establish structural guesses. This boils down to identifying introduction-elimination pairs of syntactic constructions:
\begin{minted}[mathescape]{haskell}
instance HigherOrderUnifiable TermF where
  guessMetas term = case term of
    AppF f arg -> AppF (f, [LamF ()]) (arg, [])
    -- ^ $\metavar{m}{}[\overline{z}]\;t$ implies $\metavar{m}{}[\overline{z}] := \lambda x. \metavar{m}{'}[x, \overline{z}]$
    FirstF  t  -> FirstF  (t, [PairF () ()])
    -- ^ $\pi_1\;\metavar{m}{}[\overline{z}]$ implies $\metavar{m}{}[\overline{z}] := \langle \metavar{m}{_1}[\overline{z}], \metavar{m}{_2}[\overline{z}] \rangle$
    SecondF t  -> SecondF (t, [PairF () ()])
    -- ^ $\pi_2\;\metavar{m}{}[\overline{z}]$ implies $\metavar{m}{}[\overline{z}] := \langle \metavar{m}{_1}[\overline{z}], \metavar{m}{_2}[\overline{z}] \rangle$
    _ -> bimap (,[]) (,[]) term

  shapes = [ AppF HasHead NoHead
           , FirstF HasHead, SecondF HasHead ]
\end{minted}

As we mention in Section~\ref{subsection:higher-order-matching}, the \haskell{HigherOrderUnifiable} instance can be automated entirely using either Template Haskell or GHC Generics given \haskell{Reducible} instance for the underlying bifunctor.

Finally, for type inference we specify relationships between terms and types:
\begin{minted}{haskell}
instance Inferable t
  inferF term = case term of
\end{minted}

To infer types of types, we simply need to check the types of components.
For example, for the function type we only have to check that both argument and result types are indeed types:

\begin{minted}{haskell}
    FunF inferA inferB -> do
      a <- inferA >>= shouldHaveType Universe
      b <- inferB >>= shouldHaveType Universe
      pure (TyF (FunF a b) Universe)
\end{minted}

Inferring the type of a lambda abstraction requires checking the type annotation if it is exists, inferring the type of the body, and producing the final function type:

\begin{minted}{haskell}
    LamF minferA inferBody -> do
      typeOfArg <- case minferA of
        Just inferA ->
          inferA >>= shouldHaveType Universe
        Nothing -> freshTypeMetaVar
      typedBody <- inScope typeOfArg inferBody
      typeOfBody <-
        typeOfScope typeOfArg typedBody >>= nonDep
      pure $ TyF
        (LamF (typeOfArg <$ minferA) typedBody)
        (SomeType
          (FunT Universe typeOfArg typeOfBody))
\end{minted}

Note the use of \haskell{nonDep}~--- we have to explicitly limit the inference to make sure that the type of the body does not depend on the variable bound by the lambda abstraction.

For an application term $f\;x$, we have to infer the type $F$ of the function $F$ and the type $X$ of its argument $x$. Then, if the function type $F \equiv A \to B$, then we simply need to unify argument type $X$ with the expected type $A$. Otherwise, we need to unify the type of function $F$ with type $X \to M$, where $M$ is a fresh type meta variable:

\begin{minted}[mathescape]{haskell}
    AppF inferFun inferArg -> do
      f <- inferFun -- $f : F$
      x <- inferArg -- $x : X$
      typeOfApp <- do
        typeOfFun <- typeOf f
        case typeOfFun of
          -- if $F \equiv A \to B$
          FunT _ expected result -> do
            -- then $X \equiv A$
            shouldHaveType (SomeType expected) x
            return result
          _ -> do -- otherwise
            result <- freshTypeMetaVar  -- $M : \mathcal{U_\infty}$
            argType <- typeOf x
            -- $F \equiv X \to M$
            unifyWithExpected typeOfFun
              (mkFun argType result)
            result
      return (TyF (AppF f x) (SomeType typeOfApp))
\end{minted}

Completing \haskell{inferF} for the rest of syntactic constructors in \haskell{TermF} is straightforward, and we omit the implementation to save space.
After all the preparation we get type inference for simply typed lambda calculus:
\begin{minted}[mathescape]{haskell}
> t = LamE (LamE (Var (S Z))) -- $\lambda x. \lambda y. y$
> infer t
LamT (SomeType
        (FunT Universe (MetaAppT Universe 1 [])
          (FunT Universe (MetaAppT Universe 2 [])
            (MetaAppT Universe 2 []))))
  (LamT (SomeType
          (FunT Universe (MetaAppT Universe 2 [])
            (MetaAppT Universe 2 [])))
    (Var (S Z)))
\end{minted}

The result above corresponds to the following typed term:
\begin{align}
&\lambda x. (\lambda y. y : (\metavar{m}{_2}[] : \mathcal{U_\infty}) \to (\metavar{m}{_2}[] : \mathcal{U_\infty}) : \mathcal{U_\infty}) \notag \\
&\quad : \metavar{m}{_1}[] \to ((\metavar{m}{_2}[] : \mathcal{U_\infty}) \to (\metavar{m}{_2}[] : \mathcal{U_\infty}) : \mathcal{U_\infty}) : \mathcal{U_\infty}
\end{align}

Or, omitting the $\mathcal{U_\infty}$ annotations, we get:
\begin{align}
\lambda x. (\lambda y. y : \metavar{m}{_2}[] \to \metavar{m}{_2}[]) : \metavar{m}{_1}[] \to (\metavar{m}{_2}[] \to \metavar{m}{_2}[])
\end{align}

Since we mix terms and types of STLC and use dependent type inference engine, our version of STLC has a couple of unique features, differentiating it from a classical STLC:
\begin{enumerate}
  \item We explicitly prevent the type of body in a lambda abstraction to depend on the argument. For users of STLC this means that they can input terms like $\lambda A. \lambda (x : A). x$ and get a type error saying that the inferred type of $\lambda (x : A). x$, which is $A \to A$ is dependent on the bound variable $A$, which is not allowed in STLC.
  \item We do not forbid computation in types. Indeed, a term $\lambda (f : ((\lambda x. x) A) \to B). f x$ is valid, and we can infer its type to be $(A \to B) \to B$, computing $(\lambda x. x) A \equiv A$ in the process. It is possible to add validation pass to ensure that types only consist of certain syntactic constructions, disallowing non-type terms. However, we see the ability to perform computation in types as a bonus feature for our implementation of STLC.
\end{enumerate}

Overall, we had to write down definitions of \haskell{TermF}, implement WHNF reduction for STLC terms in \haskell{Reducible} and specify how to infer types in \haskell{Inferable}. Everything else could be generated automatically with Template Haskell or GHC Generics. For this fairly little effort we have gotten an implementation of a variation of STLC with type inference and computation available in types.

\subsection{Martin-L\"of Type Theory}

Let us now apply the approach to an actual dependent type theory~--- intensional Martin-L\"of Type Theory (MLTT). We start with a generating bifunctor:
\begin{minted}[mathescape]{haskell}
data TermF scope term
  = UniverseF         -- $\text{Universe type:\;\;} \mathcal{U}$
  | PiF term scope    -- $\text{Dependent product\;\;} \Pi_{x:T_1} T_2$
  | LamF scope        -- $\text{Abstraction:\;\;} \lambda x. T_2$
  | AppF term term    -- $\text{Application:\;\;} (T_1\;T_2)$
  | SigmaF term scope -- $\text{Dependent sum\;\;} \Sigma_{x : T_1} T_2$
  | PairF term term   -- $\text{Pair:\;\;} \langle T_1, T_2 \rangle$
  | FirstF term       -- $\text{First projection:\;\;} \pi_1\;T$
  | SecondF term      -- $\text{Second projection:\;\;} \pi_2\;T$
  | IdTypeF term term -- $\text{Identity type:\;\;} x = y$
  | ReflF term        -- $\text{Reflexivity:\;\;} \mathsf{refl}_T$
  | JF term term term term term term
    -- ^ $\text{Identity type eliminator:\;\;} \mathsf{J}(A, a, C, d, x, p)$

-- | An MLTT term with free variables in a.
type Term a = FS TermF a
\end{minted}

\begin{remark}
Note that in this representation we chose to not have any type annotations for bound variables in abstraction and for the type of terms in the identity type or $\mathsf{refl}_t$. We also note that it might be possible to avoid the term $t$ in the annotation for $\mathsf{refl}_t$ as well, since the term $t$ is present in the type $t = t$ of $\mathsf{refl}$ and can be inferred in principle.
\end{remark}

In this implementation we use a single universe type and assume type-in-type: $\mathcal{U} : \mathcal{U}$. It is possible to introduce a hierarchy of universes $\mathcal{U}_0 : \mathcal{U}_1 : \mathcal{U}_2 : \ldots$ instead by using \haskell{UniverseF Natural} constructor.

Similarly to STLC implementation, we expect the relevant pattern synonyms to be written out in a mechanical way or derived automatically. Implementing WHNF reduction for MLTT is straightforward as it only differs from STLC in the use of $\mathsf{J}$-eliminator:
\begin{minted}{haskell}
instance Reducible TermF where
  reduceL = \case
    JF tA a tC d x p ->
      case reduce p of
        Refl{} -> reduce d
        p'     -> J tA a tC d x p'
    ...
\end{minted}

For \haskell{Unifiable} and \haskell{HigherOrderUnifiable} we also rely on a mechanical or automatic derivation and so omit it here to save space. Finally, we define inference for individual syntactic constructions:
\begin{minted}{haskell}
instance Inferable t
  inferF term = case term of
\end{minted}

To avoid infinite type annotations, we set the type of universe to be $\mathcal{U}_\infty$:

\begin{minted}{haskell}
    UniverseF -> pure (TyF UniverseF Universe)
\end{minted}

Inferring types for $\Pi$-types and $\Sigma$-types involves dependent type checking. Given term $\Pi_{x:A} B$, where $B$ is a subterm that may refer to $x$, we have to check that both $A : \mathcal{U}$ and $B : \mathcal{U}$. Note that since $B$ is in the scope, its inferred type, by default, might also be dependent on $x$. For example, in the term $\Pi_{x : A} \mathsf{refl}_x$ the algorithm would infer that $\mathsf{refl}_{x}$ has type $x = x$, which captures the variable $x$. To make sure the body of a $\Pi$-type is always a type, we need to unify it with $\mathcal{U}$. But for that we also need to make sure it is not dependent, so we use \haskell{nonDep}:

\begin{minted}{haskell}
    PiF inferA inferB -> do
      a <- inferA >>= shouldHaveType Universe
      typeOfA <- typeOf a
      b <- inScope typeOfA inferB
      typeOfB <- typeOfScope typeOfA b >>= nonDep
      typeOfB `shouldHaveType` Universe
      pure (TyF (PiF a b) Universe)
\end{minted}

Inferring the type for a dependent $\lambda$-abstraction is relatively straightforward.
We generate a fresh type meta variable for the argument and infer the type of the body.
In general, we should check that the inferred type is indeed a type, as many type theories, such as cubical type theory, have multiple universes. That said, in pure MLTT we can omit this check.

\begin{minted}{haskell}
    LamF inferBody -> do
      a <- freshTypeMetaVar
      typedBody <- inScope a inferBody
      b <- typeOfScope a typedBody
      typeOfScope a b >>= nonDep
        >>= shouldHaveType Universe
      pure $ TyF
        (LamF typedBody)
        (SomeType (PiT Universe a b))
\end{minted}

The rest of syntactic constructors is fairly straightforward to handle similarly.
Completing \haskell{Inferable} brings dependent type inference to MLTT.

\end{document}